\documentclass[12pt]{article}
\usepackage{amsfonts}
\usepackage{latexsym}
\usepackage{amsmath}
\usepackage{amssymb}
\usepackage{amssymb}
\usepackage[utf8x]{inputenc}

\title{The BV Formalization of Chern-Simons Theory on Deformed Superspace}
\author{Mir Faizal \\ Mathematical Institute
\\ University of Oxford\\ 
Oxford
OX1 3LB}

\begin{document}

\maketitle

\begin{abstract}
 In this paper  we will study  non-abelian Chern-Simons theory on  a deformed superspace. 
We will deform the superspace in such a way that it  includes the  noncommutativity
between bosonic  and fermionic coordinates. 
 We will first analyse the BRST and the
 anti-BRST symmetries of the Chern-Simons   theory on this deformed superspace. Then we 
will analyse  the extended BRST and the extended 
anti-BRST symmetries of this theory  in the Batalin-Vilkovisky    (BV) formalism. 
Finally, we will express these extended  BRST and  extended 
anti-BRST symmetries in 
extended superspace formalism by introducing new  Grassmann coordinates.  
\end{abstract}


\section{Introduction}

Highly supersymmetric Chern-Simons  theories are important 
because they are thought to describe the world-volume of $M2$-membranes in 
$M$-theory, at low energies \cite{1}-\cite{5}.  
 In fact, the  world-volume of $M2$-membranes in 
$M$-theory,  at low energies, is thought to be described by 
a superconformal Chern-Simons-matter theory with  with manifest $\mathcal{N} = 8$ supersymmetry \cite{ba1}. 
 A Lie $3$-algebra has been  used 
to construct an   action  for this theory  \cite{2,3}.
 However, only one example of 
such a 3-algebra  exists and  so far the   rank of the gauge group has not been increased. 

 A  $U(N) \times U(N)$
 superconformal Chern-Simons-matter  theory with level $k$ and $-k$  is also thought 
to describe the 
the world-volume of $N$ $M2$-membranes placed at the singularity of $R^8/Z_k$ \cite{apjm}.
 This theory allows arbitrary rank and but only  has  
$\mathcal{N} = 6$ supersymmetry. However, by utilizing monopole operators this symmetry gets 
enhanced to $\mathcal{N} = 8$ supersymmetry. Furthermore, after the enhancement of the supersymmetry  
this theory possess a $SO(8)$ $R$-symmetry at Chern-Simons levels
$k = 1,2$ \cite{su}. 

Chern-Simons theory in $\mathcal{N} = 1$ superfield formalism  has also been used in 
analysing the action of  $M2$-membranes \cite{14}. 
This Chern-Simons theory in $\mathcal{N}=1$ superfield formalism reproduces  
  the full Bagger and Lambert  theory \cite{2,3}
 for a $3$-algebra with totally antisymmetric structure constants.
 In doing so octonionic self-dual
tensors  are used in the construction of the real super-potential. 
The  super-potential obtained  in this way is 
only manifestly $SO(7)$ invariant. However, for specially chosen couplings, the component
action coincides with the Bagger and Lambert  action \cite{2,3}, 
and hence full $SO(8)$ symmetry is restored. Thus,
although the octonionic self-dual tensors are only $SO(7)$ invariant,
 they can be made $SO(8)$ invariant by  a special choice of
the parameters in this particular action.

 Chern-Simons theory with $\mathcal{N} =1$ supersymmetry 
 has also been studied in relation to axion gauge symmetry \cite{12}. This  occurs
in the supergravity theories arising from flux compactifications of superstrings. This  also occurs in 
the Scherk-Schwarz 
generalized dimensional 
reduction in $M$-theory. In these theories the mass term arises through a Higgs mechanism and 
the supergravity description corresponds to the gauging of some axion symmetries. These  
 are
related to shifts of the scalar fields coming from wrapped $RR$ forms
or from the $NS$ two-form $B$ field in type $II$ strings, and from the wrapped
three-form in $M$-Theory.
Chern-Simons theory with $\mathcal{N} =1$ supersymmetry coupled to parity-preserving matter fields 
has also been  
analysed using the Parkes-Siegel formulation \cite{11}.

It is expected that string theory may introduce noncommutativity in spacetime and so 
field theories with spacetime noncommutativity have been thoroughly studied \cite{dfr}-\cite{av}.  
Noncommutative tori have also been used to 
study  compactification in $M$-theory 
\cite{cds}. Noncommutativity also arises by
viewing  $M$-theory as  the $N \rightarrow\infty$ limit
 of the supersymmetric matrix
quantum mechanics describing $D0$-branes \cite{bfss}.  Furthermore, 
the  $NS$ antisymmetric tensor background
is a source of spacetime noncommutativity in string theory
 \cite{sw, dn}.

The extension of spacetime noncommutativity to superspace noncommutativity
is related to the presence of other background fields. The
$RR$ field strength background  give rise to 
$\theta$-$\theta$ type deformations \cite{ov,se}.
 Furthermore,  $x$-$\theta$ deformation is  caused by  a gravitino background \cite{bgn}.
Thus, superspace  noncommutativity  also arises in string theory. So, 
field theories with superspace noncommutativity have also been thoroughly studied \cite{se}-\cite{bs}

The BRST symmetry for the Chern-Simons theory has also been investigated
\cite{16,17}. The BRST symmetry of $\mathcal{N} = 1$  abelian   Chern-Simons theory \cite{18} 
and $\mathcal{N} = 1$ non-abelian
Chern-Simons theory \cite{19} has been discussed in the superspace formalism. 
These theories can also be analysed in the background field method. 
In this method,  all the fields in a theory  are shifted.  The BRST 
and the anti-BRST symmetries of these shifted fields can then be  analysed  in the
Batalin-Vilkovisky   formalism \cite{10a}-\cite{11a}.  
In this formalism the extended BRST and the extended anti-BRST 
symmetries arise  due to the
 invariance of a
 theory under both the original BRST and the original anti-BRST transformations along with 
these shift transformation. This  
    has been done for the conventional 
Yang-Mills theories and the conventional  Chern-Simons theories \cite{12a}-\cite{13a}. 
Furthermore, the extended BRST and the extended anti-BRST  symmetries 
have been analysed in the extended 
superspace formalism~\cite{14a, 30}. This is because 
the BRST and the anti-BRST symmetries mix the fermionic and bosonic 
coordinates and can thus be viewed as  supersymmetric transformations. 

In this paper we will analyse the  $\mathcal{N} =1$ Chern-Simons 
  theory in a  deformed superspace. 
The deformation of the superspace 
 will break the supersymmetry of the theory. Then we 
will analyse the extended BRST and the extended 
 anti-BRST symmetries for this 
Chern-Simons theory in the BV-formalism. 
Finally we shall express our results in an extended superspace 
formulation by introducing new Grassmann coordinates. 
 It may be noted even though
the results of this paper can easily  be first generalized and then 
used  to analyse the deformation of the ABJM theory, 
this will not be done here. In fact, in this paper, we will only analyse the 
pure Chern-Simons   theory
with no matter fields.

\section{  Deformed Super-Chern-Simons Theory  }
In this section we shall construct a three dimensional   Chern-Simons   theory   
on a deformed superspace. 
To do so we define $\theta^a$ as a two-component Grassmann parameter and let 
$ y^\mu = x^\mu + \theta^a (\gamma^\mu) _{a}^b \theta_b$.  
Then we promote them to operators $\hat{ \theta}^a$ and $\hat{y}^\mu$ and impose the following 
 deformation of the superspace algebra, 
\begin{eqnarray}
 \{ \hat{\theta}^a, \hat{\theta}^b \} = C^{ab}, && [\hat{y}^\mu, \hat{y}^\nu] = 
 B^{\mu\nu}, \nonumber \\ 
 {[\hat{y}^\mu, \hat{\theta}^a]} = A^{\mu a }. && 
\end{eqnarray} 
We use Weyl
ordering and  express the Fourier transformation of this superfield as, 
\begin{equation}
\hat{\Gamma}_a (\hat{y}, \hat{\theta}) =
\int d^3 k \int d^2 \pi e^{-i k \hat{y} -\pi \hat{\theta} } \;
\Gamma_a (k,\pi).
\end{equation}
Now  we  have a one to one map between a function of
$\hat{\theta}, \hat{y}$ to a function of ordinary
 superspace coordinates $\theta, y$ via
\begin{equation}
\Gamma_a (y, \theta)  =
\int d^3 k \int d^2 \pi e^{-i k y -\pi \theta } \;
\Gamma_a (k,\pi),
\end{equation}
where 
\begin{equation}
 \Gamma_a (y, \theta) = \chi_a + B\theta_a + \frac{1}{2} ( \gamma^\mu \theta A_\mu)_a  + i \theta^2 \left[\lambda_a -
 (\frac{1}{2 }\gamma^\mu \partial_\mu \chi)_a\right].
\end{equation}
 We can express the product of two fields  
${\hat{\Gamma}^a}(\hat{y},\hat{\theta}) { \hat{\Gamma}_{a  }} (\hat{y},\hat{\theta})$
on this deformed superspace as
\begin{eqnarray}
{\hat{\Gamma}^a}(\hat{y},\hat{\theta}) { \hat{\Gamma}_{a  }}  (\hat{y},\hat{\theta}) &=&
\int d^3 k_1 d^3 k_2 \int d^2 \pi_1  d^2 \pi_2
\exp -i( ( k_1 +k_2) \hat{y} +(\pi_1 +\pi_2) \hat{\theta} )  \nonumber \\ && 
\,\,\,\,\,\, \,\,\,\,\,\, \,\,\,\,\,\, \,\,\,\, \,\,\,\,\,\, \,\,\,\, \,\,\,\,\,\, \,\,\,\, 
\,\,\,\,\,\, \times  \exp(i\Delta)
{\Gamma}^a  (k_1,\pi_1) {\Gamma}_{a  }   (k_2,\pi_2),
\end{eqnarray}
where
\begin{equation}
\exp (i\Delta) = \exp -\frac{i}{2} \left(
C^{ab}\pi^1_a  \pi^2_b
+ B^{\mu\nu} k^2_\mu k^1_\nu
+  A^{\mu a} (\pi^2 _a k^1_\mu - k^2_\mu \pi^1_a \right),
\end{equation}
So we can now define the star product  between ordinary functions  as follows:
\begin{eqnarray}
{\Gamma^a}(y,\theta) \star { \Gamma_{a  }}  (y,\theta) & =& 
\exp -\frac{i}{2} \left(
C^{ab}\partial^1_a  \partial^2_b
+ B^{\mu\nu} \partial^2_\mu \partial^1_\nu
+ A^{\mu a} (\partial^2 _a \partial^1_\mu - \partial^2_\mu \partial^1_a \right)) \nonumber \\ &&
\,\,\,\,\,\, \,\,\,\,\,\, \,\,\,\,\,\,\,\,\,\,
 \times 
 {\Gamma^a}(y_1,\theta_1) { \Gamma_{a  }}  (y_2, \theta_2)
\left. \right|_{y_1=y_2=y, \; \theta_1=\theta_2=\theta}.
\label{star2}
\end{eqnarray}
The star product reduces to the usual Moyal star product
for the bosonic noncommutativity  in the limit $C^{ab}=A^{a\mu}=0$ and 
for $A^{a\mu}=C^{ab}=0$ it reduces to the standard 
fermionic star product. It is also useful to define the following bracket 
\begin{equation}
 [\Gamma^a \star \Gamma_a] = \frac{1}{2} T_A f^A_{BC} \Gamma^{a B} \star \Gamma_a^C.
\end{equation}

In order to construct a  Chern-Simons  
 theory on this deformed superspace,
 it is useful to define the following fields  
\begin{eqnarray}
 \Omega_a & = & \omega_a - \frac{1}{6}[\Gamma^b\star \Gamma_{ab}], \\
 \omega_a & = & \frac{1}{2} D^b D_a \Gamma_b 
- \frac{i}{2}  [\Gamma^b \star D_b \Gamma_a] -
 \frac{1}{6} [ \Gamma^b \star
[ \Gamma_b \star \Gamma_a] ], \label{omega} \\
 \Gamma_{ab} & = & -\frac{i}{2}  \left(D_{(a}\Gamma_{b)} 
- i[\Gamma_a \star \Gamma_b]\right),
\end{eqnarray}
where 
\begin{equation}
D_a = \partial_a + (\gamma^\mu \theta \partial_\mu)_a.
\end{equation}
 Now the non-abelian    Chern-Simons   theory  on this deformed superspace  can now be
 written as 
\begin{equation}
 \mathcal{L}_c = \int d^2 \,  \theta \, \,  
 Tr (  \Gamma^a  \star  \Omega_a) _|.
\end{equation}
and $'|'$  means that the quantity is evaluated at $\theta_a =0$. We consider our theory
 to be defined on a manifold without 
a boundary. This theory on an undeformed superspace 
has $\mathcal{N} =1$ supersymmetry, 
however on  the deformed superspace the
  supersymmetry is broken.  This is because of the deformation $ \{ \hat{\theta}^a, \hat{\theta}^b \} = C^{ab}$, 
the supersymmetry corresponding to $Q_a$ is broken \cite{se}-\cite{bs}. 
Here $Q_a$ is the generator of $\mathcal{N}=1$ supersymmetry and is given by 
\begin{equation}
 Q_a = \partial_a - (\gamma^\mu \theta \partial_\mu)_a.
\end{equation}
For theories with $\mathcal{N}=2$ supersymmetry in three dimensions  we have two super-charges. It is thus possible to deform the super-algebra 
corresponding to 
one of them and leave super-algebra corresponding to the other one undeformed. 
So we, can break the supersymmetry from $\mathcal{N} =2$ supersymmetry to $\mathcal{N} =1$ supersymmetry by such deformations. 
In four dimensions, we can also break the supersymmetry to $\mathcal{N}=1/2$ supersymmetry. 
However, for $\mathcal{N}=1$ supersymmetry in three dimensions such a deformation breaks all the supersymmetry of the theory.

All the degrees of freedom in this Lagrangian density are not physical as 
it is invariant under the following  super-gauge transformations 
\begin{equation}
 \delta \Gamma_a = (D_a -i \Gamma_a) \star \Lambda, \label{sgt}
\end{equation}
where $\Lambda = \Lambda^A T_A $. 
 In order to quantise this theory we will have to add a gauge fixing term and a ghost term to it. 
We choose the gauge fixing term $\mathcal{L}_{gf}$ and the ghost term $\mathcal{L}_{gh}$ as follows:
\begin{eqnarray}
\mathcal{L}_{gf} &=& \int d^2 \theta \, \,    \, Tr (F  \star D^a \Gamma_a )_|, \nonumber \\
\mathcal{L}_{gh} &=&  \int d^2 \theta \, \,    \, Tr (\overline{c}  D^a  (D_a -i \Gamma_a)  \star c)_|. 
\end{eqnarray}
Here $B$ is a matrix  valued scalar superfield,  $c$ and $\overline{c}$ are matrix
 valued anti-commutating superfields. These superfields are 
 suitably contracted with generators of the Lie algebra in the adjoint representation, 
\begin{eqnarray}
 F (y, \theta) = F^A(y, \theta) T_A, &&
c(y, \theta) =c^A(y, \theta) T_A, \nonumber \\ 
\overline{c}(y, \theta) = \overline{c}^A(y, \theta) T_A.&&
\end{eqnarray}
The total Lagrangian density which is obtained by the sum of the original Lagrangian density with 
the gauge fixing term and the ghost term  is invariant under the following BRST transformations
\begin{eqnarray}
s \Gamma_a=   (D_a -i \Gamma_a) \star  c, &&
s c = -\frac{1}{2} [c\star c],     \\ \nonumber
s \overline{c}=-   F, &&
s F=0,
\end{eqnarray}
where 
\begin{equation}
 [c\star c] = \frac{1}{2} T_A f^A_{BC}  c^B \star c^C.
\end{equation}
In fact, this total Lagrangian density  is also invariant under the following anti-BRST transformations 
\begin{eqnarray}
\overline s \Gamma_a =   (D_a -i \Gamma_a)   \star \overline{c}, &&
\overline s c = F- [c \star \overline c], \\ \nonumber
\overline s \overline c = -\frac{1}{2}   [ \overline{c}\star \overline{c}], &&
\overline s F= [F \star \overline c].
\end{eqnarray}
These transformations satisfy 
\begin{equation}
 s^2 = \overline s ^2 =0.
\end{equation}
In fact, they also satisfy $s \overline s + \overline s s =0$. 
Now the sum of the gauge fixing term  and the ghost term  can be written as 
\begin{eqnarray}
\mathcal{L}_{gf} + \mathcal{L}_{gh} & =& -  \int d^2 \theta \, \,  
  \frac{s \overline s }{2} Tr (\Gamma^a \star \Gamma_a)_| \nonumber \\ 
 & =& \int d^2 \theta \, \,    \frac{\overline{s} s }{2} Tr (\Gamma^a \star \Gamma_a)_|.
\end{eqnarray}
The  BRST and the anti-BRST invariance of the total 
 Lagrangian density given by 
\begin{equation}
 \mathcal{L} = \mathcal{L}_c  +  \mathcal{L}_{gf} + \mathcal{L}_{gh},
\end{equation}
follows from the fact that these BRST and the anti-BRST transformations 
are nilpotent and the sum of the gauge fixing term and the ghost term can be written as a total
 BRST and  a total anti-BRST variation. 
For the original classical Lagrangian density the BRST and anti-BRST transformations are just 
gauge transformations with $\Lambda$ replaced by the ghosts or the
anti-ghosts. 
Thus, the total Lagrangian density is invariant  under the BRST and the anti-BRST transformations. 

\section{Extended BRST}

In this section we will analyse  the extended BRST invariant Lagrangian density. To do so we first
 shift the original fields as
\begin{eqnarray}
 \Gamma_a &\to&  \Gamma_a - \tilde{ \Gamma }_a, \nonumber \\
c &\to& c - \tilde{c}, \nonumber \\
\overline{c} &\to& \overline{c} - \tilde{\overline{c}},\nonumber\\
F &\to&  F - \tilde{ F}.
\end{eqnarray}
The extended BRST invariant Lagrangian density is obtained by
 requiring it to be invariant under both the original BRST transformations and these shift
transformations of the original fields
\begin{equation}
\tilde{\mathcal{ L}}=\mathcal{L}(\Gamma_a - 
\overline {\tilde \Gamma_a}, c- \overline{\tilde c}, \overline{c}-\overline{\tilde c}, F- \tilde F).
\label{9a}
\end{equation}
In order to discuss the extended BRST invariant Lagrangian 
density it will be useful to 
define $\tilde{\nabla}_a$ as 
\begin{equation}
 \tilde{\nabla}_a = D_a -i \Gamma_a + i \tilde{\Gamma}_a. 
\end{equation}
Now the extended BRST invariant Lagrangian density  is invariant under the following  
extended BRST  transformations
\begin{eqnarray}
s \Gamma_a= \psi_{a}, &&
s \tilde \Gamma_a= \psi_{a} -\tilde{\nabla}_a \star (c- \tilde c),\\ \nonumber
s c =   \epsilon &&  s \tilde c=  \epsilon + \frac{1}{2}[(c- \tilde c) \star ( c- \tilde c)],\\ 
\nonumber
s \overline{c} =    \overline{\epsilon},  &&
s \tilde{\overline{c}}= \overline{\epsilon} + (F-\overline{F}), \nonumber \\ 
s F =\psi,  &&
s \tilde F=\psi,
\end{eqnarray}
where 
\begin{equation}
 \tilde{\nabla}_a \star (c- \tilde c) = D_a \star (c- \tilde c)  - i (\Gamma_a - \tilde{\Gamma}_a)
 \star (c- \tilde c).
\end{equation}
Here $\psi_a, \epsilon, \overline{\epsilon} $ and $\psi$ are the ghost fields associated with the 
shift symmetries of the original fields
 $ \Gamma_a, c,
 \overline{c}$ and $F$ respectively.
The   BRST transformations of these ghosts associated with the shift symmetry  vanish, 
\begin{eqnarray}
s\psi_{a} =0, && 
s\epsilon =0,\\ \nonumber
s \overline{\epsilon}=0, &&
s \psi=0,
\end{eqnarray}
We  add anti-fields with opposite parity to the original fields. These  anti-fields transform into new
auxiliary fields  $b_a, B,  \overline{B}$ and $b $ under BRST transformations,
\begin{eqnarray}
s \Gamma^{*}_{a}=-   b_a, &&  
s c^{*}=- B,\\ \nonumber
s \overline c^{*}=-   \overline{B}, && 
s F^{*}=-   b.
\end{eqnarray}
The BRST transformations of these new auxiliary fields also vanish
\begin{eqnarray}
s b_a=0, &&
s B=0,\nonumber \\
s \overline{B}=0, &&
s b=0.\label{3b}
\end{eqnarray}
We now choose the  Lagrangian density to gauge fix the shift symmetry in such a way that the 
tilde fields will be made to
 vanish so that we can recover the original theory
\begin{eqnarray}
 \tilde {\mathcal{L}}_{gf}+  \tilde {\mathcal{L}}_{gh} = \int d^2 \theta \, \, 
  Tr(-b^a    \star \tilde{\Gamma}_{a} + \Gamma^{*a} \star 
(\psi_{a}-\tilde{\nabla}_a  \star (c- \tilde{c}))\nonumber \\ 
- \overline B \star  \tilde{c}+\overline{c}^{*}\star \left(\epsilon+\frac{1}{2}
[(c-\tilde{c}) \star (c-\tilde{c})]\right)\nonumber \\ 
+B \star \tilde{\overline{c}}-c^{*} \star (\overline{\epsilon}+(F-\tilde{F})) +b 
  \star \tilde{F}+F^{*}\star \psi)_|.
\end{eqnarray}
Here tilde fields  vanish upon integrating out the auxiliary fields $b_a, B, \overline{B}$ and $b$.
 This Lagrangian density is  invariant under the original BRST transformation and the shift
 transformations.
 Along with this Lagrangian density we have the original Lagrangian density, which is only a
 function of the original fields.
 So we define  $\Psi $ as
\begin{equation}
\mathcal{L}'_{gf} + \mathcal{L}'_{gh} = \int d^2 \theta \, \,  Tr (s \Psi)_|.  \label{8a}
\end{equation}
Expanding this Lagrangian density, we obtain
\begin{equation}
\mathcal{L}'_{gf} + \mathcal{L}'_{gh} =  \int d^2 \theta \, \,   
Tr\left(-\frac{\delta  \Psi}{ \delta \Gamma_a}\star 
\psi_{a}+\frac{\delta  \Psi}{ \delta c}\star \epsilon
 +\frac{\delta  \Psi}{ \delta  \overline{c}}\star \overline{\epsilon}-\frac{\delta \Psi}{\delta F}\star
 \psi\right)_|.
\end{equation}
If we integrate out the fields setting the tilde fields to zero, we have
\begin{eqnarray}
\mathcal{L}&=&\mathcal{L}_c  +\tilde {\mathcal{L}}_{gf}+ \tilde{\mathcal{L}}_{gf} + \mathcal{L}'_{gf} 
+ \mathcal{L}'_{gh} 
\nonumber  \\  &=& \mathcal{L}_c (\Gamma_a  -\tilde{\Gamma}_a ) 
+\int d^2 \theta \, \,    Tr\left(\Gamma^{*a}\star \nabla_a  c+ \frac{1}{2}\overline{c}^{*}\star
 [c  \star c]  -c^{*}\star 
F\right. \nonumber \\ && \left. -\left(\Gamma^{*a}  +\frac{\delta \Psi}{\delta \Gamma_a}\right)\star \psi^{a} 
+ \left(\overline{c}^{*} + \frac{\delta  \Psi}{ \delta c}\right)\star\epsilon-\left(c^{*} + 
\frac{\delta  \Psi}{ \delta  \overline{c}}\right)
\star\overline{\epsilon} \right. \nonumber \\ && \left. +\left(F^{*}- 
\frac{\delta \Psi}{\delta F}\right)\star \psi  \right)_|.
\end{eqnarray}
The explicit expression for the anti-fields is achieved by  integrating out the ghosts associated with the shift symmetry,
\begin{eqnarray}
 \Gamma^{*a}= -\frac{\delta\Psi}{\delta \Gamma_a},&&
\overline{c}^*= -\frac{\delta \Psi}{\delta c},\nonumber\\
c^{*}=\frac{\delta \Psi}{\delta \overline{c}},&&
F^{*}=\frac{\delta \Psi}{\delta F}.
\end{eqnarray}
With these identifications we obtain an explicit form for the Lagrangian density which is invariant under the extended BRST transformations.
\subsection{Extended Anti-BRST Lagrangian}
In the previous sections we analysed the extended BRST symmetry.
 Now we will discuss the extended anti-BRST symmetry. 
 The original Lagrangian density is also    invariant under the   following 
extended anti-BRST transformations,
\begin{eqnarray}
\overline{s} \Gamma_a&=  &  \Gamma^{*}_{a} +\tilde{\nabla}_a \star (c-\tilde{\overline{c}}), \nonumber \\ 
\overline{s}c&=   & c^{*}+(F-\overline{F})-   [(c -\tilde{c}) \star (\overline{c}-\tilde{\overline{c}})], \nonumber \\ 
\overline{s}\overline{c}&=  &  \overline{c}^{*}- \frac{1}{2}[(\overline{c}-\tilde{\overline{c}}) \star
 ( \overline{c}-\tilde{\overline{c}})], 
\nonumber \\ 
\overline{s} F &= &   F^*    + [(F - \tilde{F}) \star ( \overline{c}-\tilde{\overline{c}})],
\end{eqnarray}
and shifted super-fields have the following 
extended anti-BRST transformations,
\begin{eqnarray}
 \overline{s}\tilde{\Gamma_a}=  \Gamma^{*}_{a}, &&
 \overline{s}\tilde{c}=    c^{*},\\ \nonumber
\overline{s}\tilde{\overline{c}}=     \overline{c^{*}}, &&
 \overline{s}\tilde{F}=    F^{*}.
\end{eqnarray}
The ghost fields associated with the shift symmetry have the following extended anti-BRST transformations,
\begin{eqnarray}
\overline{s} \psi_a & =& b_a + \tilde{\nabla}_a  \star( F -\tilde{F} ) - [(\tilde{\nabla}_a \star ( c -\tilde c)) \star  
( \overline{c}-\tilde{\overline{c}})],\\ \nonumber
\overline{s} \epsilon&=&   B - [(F-\tilde{F})\star( \overline{c}-\tilde{\overline{c}})]
+[[(\overline{c}-\tilde{\overline{c}})\star(c-\tilde{c})] \star (c-\tilde{c})], \\ \nonumber
\overline{s}\overline{\epsilon}&=&    \overline{B} - [(F-\tilde{F})\star( \overline{c}-\tilde{\overline{c}})],\\ \nonumber
\overline{s} \psi&=&   b.
\end{eqnarray}
The extended anti-BRST transformations of the anti-fields and the auxiliary fields associated with the shift symmetry vanish,
\begin{eqnarray}
\overline{s }b_a =0, && \overline{s}  \Gamma ^{*}_a =0,\nonumber\\
\overline{s} B=0, &&\overline{s} c^{*}=0,\nonumber\\
\overline{s} \overline{B}=0, &&\overline{s} \overline c^{*}=0,\nonumber\\
\overline{s }\overline{b}=0, && \overline{s} F^{*}=0.
\end{eqnarray}
For the Lagrangian density, which is both BRST and anti-BRST invariant, it follows that it  must also be invariant 
under the extended anti-BRST transformation
 at least on-shell, where these transformations reduce to the original  anti-BRST transformations.

\subsection{Extended Superspace Formulation}
The BRST transformations  and the anti-BRST transformations are also  
supersymmetry transformations as they mix bosonic and fermionic fields. 
In fact the BRST and the anti-BRST transformations have been expressed in
superspace even for bosonic theories \cite{14a, 30}. 
We will now express the extended BRST and the extended anti-BRST
 invariance of this  Chern-Simons   theory in superspace. 
This superspace has nothing to do with the original superspace and a similar structure will exists even for a bosonic theory. 
It is related to the supersymmetric nature of the extended BRST and the extended anti-BRST symmetries. 
Thus we introduce two anti-commutating Grassmann parameters $\xi$ and $\overline{\xi}$ and define the following superfields with them, 
\begin{eqnarray}
 \phi_{a}(x,\xi,\overline{\xi})&=& \Gamma_a + \xi \psi_{a}+\overline{\xi}(\Gamma_a^{*}+
\tilde{\nabla}_a  \star (\overline{c}-\tilde{\overline{c}}))
 \nonumber \\
&& +\xi \overline{\xi}(b_a   +\tilde{\nabla}_a  \star (F-\tilde{F}) - [(\tilde{\nabla}_a  \star
(c-\tilde{c}))\star (\overline{c}-\tilde{\overline{c}})]),
 \nonumber \\
\tilde{\phi}_{a}(x,\xi,\overline{\xi})&=&\tilde{\Gamma}_{a}+\xi(\psi_{a}-\tilde{\nabla}_a \star(c-\tilde{c})
+\overline{\xi}\Gamma_a^{*}+\xi \overline{\xi}b_a,  
\nonumber  \\ 
\eta (x,\xi,\overline{\xi})&=&c+\xi \epsilon
+\overline{\xi}(c^{*}+(F-\tilde{F})-[(c-\tilde{c})\star(\overline{c}-\tilde{\overline{c}})] )\nonumber \\ 
&& +\xi\overline{\xi} (B-[(F-\tilde{F})\star(c-\tilde{c})] + [[(\overline{c}-\tilde{\overline{c}})
 \star(c-\overline{c})]\star(c-\overline{c})]),
\nonumber \\ 
\tilde{\eta}(x,\xi,\overline{\xi})&=&\tilde{c}+\xi(\epsilon+\frac{1}{2}[(c-\tilde{c})\star( c-\tilde{c})]) + \overline{\xi}c^{*}
+\xi\overline{\xi}B,
\nonumber \\ 
\overline{\eta}(x,\xi,\overline{\xi})&=&\overline{c}+\xi\overline{\epsilon}+\overline{\xi}(\overline{c}^{*}-
\frac{1}{2}[(\overline{c}-\tilde{\overline{c}})\star(\overline{c}-\tilde{\overline{c}})]) 
\nonumber \\ 
 && +\xi \overline{\xi}(\overline{B}-[(F-\tilde{F})\star(\overline{c}-\tilde{\overline{c}})]),
\nonumber \\ 
\tilde{\overline{\eta}}(x,\xi,\overline{\xi})&=&\tilde{\overline{c}}+\xi(\overline{\epsilon}
+(F-\tilde{F})) + \overline{\xi}\overline{c}^{*}+\xi \overline{\xi}\overline{B}.
\end{eqnarray}
Now we have 
\begin{eqnarray}
-\frac{1}{2}\int d\xi d\overline{\xi} d^2 \theta \, \,Tr (\tilde{\phi}^a \star
\tilde{\phi}_a)_| &=&  \int  d^2 Tr (- \Gamma^{*a} \star (\psi_{a}-\tilde{\nabla}_a \star (c-\tilde{c})) \nonumber \\ && 
 -b_a  \star \tilde{\Gamma}^{a} )_|,
\nonumber \\ 
\int d\xi d\overline{\xi} d^2 \theta \, \, Tr(\tilde{\eta}\star \tilde{\overline{\eta}})_|&=&
 \int  d^2 \theta Tr \left( - \overline{\tilde{B}}\tilde{c}  +B\star \tilde{\overline{c}} -\overline{B}\star \tilde{c} \right.\nonumber \\ 
 &&  + \overline{c}^{*}\star \left(\epsilon+\frac{1}{2}
[(c-\tilde{c})\star(c-\tilde{c})]\right)  \nonumber \\ 
& & \left.
 - c^{*}\star (\overline{\epsilon}+(F-\tilde{F}))\right)_|.
\end{eqnarray}
Thus the sum of the shifted gauge fixing term and the shifted ghosts term can be written as,  
\begin{eqnarray}
\tilde {\mathcal{L}}_{gf}+ \tilde{\mathcal{L}}_{gf}  &= &\int d\xi d\overline{\xi} d^2 \theta \, \,  
Tr(-\frac{1}{2}\tilde{\phi}^a\star  \tilde{\phi}_a +\tilde{\eta}\star \tilde{\overline{\eta}})_| \nonumber \\ &=&
  \int  d^2 \theta \,\, Tr (-b^a  \star  \tilde{\Gamma}_{a}-\Gamma^{*a}\star (\psi_{a} - {\nabla}_a \star 
(c-\tilde{c}))\nonumber \\  &&  -\overline{B}\star \tilde{c} 
+\overline{c}^{*}\star \left(\epsilon +\frac{1}{2}[(c-\tilde{c})\star(c-\tilde{c})]\right)\nonumber \\ 
&&+B\star \tilde{\overline{c}}-c^{*}\star (\overline{\epsilon}+(F-\tilde{F})))_|.
\end{eqnarray}
Being the $\xi \overline{\xi}$ component of a superfield, this gauge fixing Lagrangian is manifestly invariant under 
extended BRST and anti-BRST transformation.  
We now  note that we can choose a fermionic superfield such that \cite{14a}, 
\begin{equation}
\Phi(x,\xi,\overline{\xi})=\Psi+\xi s \Psi+ \overline{\xi}\overline{s}\Psi
+\xi \overline{\xi} s \overline{s}\Psi.
\end{equation}
 The $\xi \overline{\xi}$ component of this equation vanishes after  using the equations of motion. 
This is expected as the sum of the original gauge fixing term and the ghost term is
 invariant under the BRST and the anti-BRST transformations. Thus the $\xi \overline{\xi}$ component of $\Phi(x, \xi, \overline{\xi})$ 
vanish on shell because when we use the equations of motion, the tilde fields vanish and the theory reduces to the 
original theory which   is invariant under  the BRST and the anti-BRST transformations. 
We can now express the sum of the original gauge fixing term and the ghost term as, 
\begin{eqnarray}
\mathcal{L}'_{gf} + \mathcal{L}'_{gf}&=& \int d\xi d\overline{\xi} d^2 \theta\,\, 
 Tr (\delta (\overline{\xi}) \Phi(x, \xi,\overline{\xi}))_|.
\end{eqnarray}
This is not only invariant under extended BRST transformations but it is 
 also  invariant under extended anti-BRST transformations on-shell.  
  The total  Lagrangian density is invariant
 under extended BRST transformations. It is  also invariant under extended anti-BRST transformations on-shell. 
We now write this total Lagrangian density as,  
\begin{eqnarray}
\tilde{\mathcal{L}}&=& \mathcal{L}_c  +\tilde{\mathcal{L}}_{gf}+\tilde{\mathcal{L}}_{gf} + 
\mathcal{L}'_{gf}+\mathcal{L}'_{gf}\\ \nonumber
&=&\mathcal{L}_c (\Gamma_a  -\tilde{\Gamma}_a )+ \int d\xi d\overline{\xi} d^2 \theta \, \,  
Tr\left(-\frac{1}{2}\tilde{\phi}^a \star \tilde{\phi}_a +\tilde{\eta}\star \tilde{\overline{\eta}}\right)_| \\ \nonumber && + 
 \int d\xi d\overline{\xi} d^2 \theta \,\, Tr  (\delta (\overline{\xi}) \Phi(x, \xi_1,\overline{\xi}))_|.
\end{eqnarray}
As  $F$ and $\tilde{F}$ are auxiliary fields so  we   redefine them as 
$F-\tilde{F}\rightarrow F$ and    integrated out $(F+\tilde{F})$. The   constant thus obtained can be 
  absorbed into the normalisation constant.
 
\subsection{Conclusion}
In this paper we analysed the 
Chern-Simons theory in deformed superspace where 
the deformation included the noncommutativity
between bosonic coordinates and fermionic coordinates. 
We  found   that the sum of the original 
  classical Lagrangian density, a gauge fixing term and a ghost term was invariant
under the the BRST and the anti-BRST transformations.  
 We  also analysed the extended BRST and the extended anti-BRST 
symmetries of this theory in the extended superspace formalism. 
 This theory  was found to be invariant under extended BRST transformations.
 It was also found to be invariant   under  on-shell
extended anti-BRST  transformations.

The spacelike noncommutative field theories are known to be unitarity \cite{u}-\cite{u2}.  
However,  due to Eq. $(\ref{star2})$,
 infinite temporal derivatives will occur in the product of fields 
for  field theories with spacetime noncommutativity. 
It is well known that
the evolution of the $S$-matrix is not unitary  for the 
field theories with higher order temporal derivatives \cite{3an}-\cite{14n}.
Thus, spacetime noncommutativity  will   break the unitarity of the resultant theory. 
However, if we restrict the theory to spacelike noncommutativity and thus do not include any 
higher order temporal derivatives then this problem can be avoided. 
In fact, in the case of spacelike noncommutativity it is possible 
to construct the Norther's charges \cite{nm1}-\cite{nm2}.
Thus, if we restrict the spacetime deformations to spacelike noncommutativity then 
 we can construct the Norther's charges for this deformed theory.
 It will be interesting to construct the BRST and the anti-BRST charges for this theory 
and use them to find the physical states in this theory.  

It will be interesting to generalise the result of 
this paper to Chern-Simons theories with higher supersymmetry, coupled to matter fields.
 In particular, the analysis of  
 $U(N) \times U(N)$ Chern-Simons gauge theory, with level $k$ and $-k$  
enhanced to $\mathcal{N} =8$ supersymmetry for $k =1,2$, suitably coupled to matter fields
 in this deformed superspace will have important consequences for $M$-theory.
Due to the duality between $M$-theory and $II$ string theory, we expect that a 
noncommutative deformation of the super algebra on the string theory side will 
correspond to some deformation of the super algebra on the $M$-theory side.  
In fact, just like a background two-form  field 
strength becomes a sources of noncommutativity for 
$D$-branes, a background three-form field 
suitably coupled to the ABJM theory could also lead to the 
noncommutativity. It may be noted that a three-form field strength occurs 
naturally in $M5$-branes. Furthermore, $M5$-branes in $M$-theory act as analogous objects 
to a $D$-brane in string theory, in the sense that $M2$-branes  can end on them.
Thus, the coupling of ABJM theory to a background three-form field strength 
can be useful in describing the physics of $M2$-branes ending on $M5$-branes.  
 As the action for a single 
$M5$-brane  can be derived by demanding
the $\kappa$-symmetry of the open membrane ending on it \cite{14om}, the analysis of 
ABJM theory coupled to a background three-form field strength might 
give some useful insights into understanding the dynamics of multiple $M5$-branes.
It may be noted that even though the action for a single $M5$-brane is known, the  
action   for multiple $M5$-branes is not
known \cite{41s}-\cite{42s}.
 Coupling of the ABJM theory to other background fields could 
lead to other superspace deformations of the super algebra. So, it will also be interesting 
to analyse the consequences of  coupling of ABJM theory to other background fields.   

Chern-Simons theories also  have important  applications in  condensed matter physics. 
This is because of their  relevance 
 to the fractional quantum Hall effect, which is based 
on the concept of statistical transmutation. 
\cite{a}-\cite{d}. 
In two dimensions, fermions can be described as charged bosons carrying
an odd integer number of  flux quanta. This is achieved by  analysing  
 Chern-Simons fields coupled to these bosons. Then, the electrons in an external
magnetic field can be described as bosons in a combined external and statistical magnetic
field. At special values of the filling fraction the statistical 
field cancels the external field,
in the mean field sense. At these values of 
the filling fraction and the system is described as a gas of bosons feeling no net
magnetic field. Thus, these bosons condense into a homogeneous ground state. 
This model also describes the  existence of vortex and anti-vortex excitations.

Lately, supersymmetric 
generalisation of the fractional quantum Hall effect has also been investigated \cite{a11}-\cite{a4}.
In particular, physical properties of the topological
excitations in the supersymmetric quantum Hall liquid have been discussed using a dual supersymmetric
Chern-Simons theory \cite{a5}.  Furthermore, the fractional quantum Hall 
effect is closely related to  noncommutativity  of the spacetime \cite{fqhfqh1}-\cite{fqhfqh2}. 
Thus, the results of this paper  can have interesting condensed matter applications. 
This is because we can analyse the superspace deformation 
of the supersymmetric fractional quantum Hall effect. It can
  change the behavior of  fractional condensates and thus
have  important consequences for the transport properties in the quantum hall  system.
 Holography has also been used to analyse the  supersymmetric
 fractional quantum Hall effect \cite{e}. 
In fact, supersymmetric
Chern-Simons theories have   been used to study various interesting
 examples of $AdS_4/CFT_3$ correspondence \cite{6}-\cite{10}.  
 It will be interesting  to
 analyse similar effects in the deformed superspace theories
 with superspace noncommutativity.

\end{document}